\input harvmac

\chardef\tempcat=\the\catcode`\@
\catcode`\@=11
\def\cyracc{\def\u##1{\if \i##1\accent"24 i%
    \else \accent"24 ##1\fi }}
\newfam\cyrfam
\font\tencyr=wncyr10
\def\cyr{\fam\cyrfam\tencyr\cyracc}


\def\unlockat{\catcode`\@=11}
\def\lockat{\catcode`\@=12}

\unlockat

\def\newsec#1{\global\advance\secno by1\message{(\the\secno. #1)}
\global\subsecno=0\global\subsubsecno=0\eqnres@t\noindent
{\bf\the\secno. #1}
\writetoca{{\secsym} {#1}}\par\nobreak\medskip\nobreak}
\global\newcount\subsecno \global\subsecno=0
\def\subsec#1{\global\advance\subsecno by1\message{(\secsym\the\subsecno. #1)}
\ifnum\lastpenalty>9000\else\bigbreak\fi\global\subsubsecno=0
\noindent{\it\secsym\the\subsecno.}{#1}
\writetoca{\string\quad {\secsym\the\subsecno.} {#1}}
\par\nobreak\medskip\nobreak}
\global\newcount\subsubsecno \global\subsubsecno=0
\def\subsubsec#1{\global\advance\subsubsecno by1
\message{(\secsym\the\subsecno.\the\subsubsecno. #1
)}
\ifnum\lastpenalty>9000\else\bigbreak\fi
\noindent\quad{\secsym\the\subsecno.\the\subsubsecno.}{#1}
\writetoca{\string\qquad{\secsym\the\subsecno.\the\subsubsecno.}{#1}}
\par\nobreak\medskip\nobreak}

\def\subsubseclab#1{\DefWarn#1\xdef #1{\noexpand\hyperref{}{subsubsection}%
{\secsym\the\subsecno.\the\subsubsecno}%
{\secsym\the\subsecno.\the\subsubsecno}}%
\writedef{#1\leftbracket#1}\wrlabeL{#1=#1}}
\lockat


\def\IC{{\bf C}}

\def\IG{\relax\hbox{$\inbar\kern-.3em{\rm G}$}}
\def\IGa{\relax\hbox{${\rm I}\kern-.18em\Gamma$}}
\def\IH{\relax{\rm I\kern-.18em H}}
\def\II{\relax{\rm I\kern-.18em I}}
\def\IK{\relax{\rm I\kern-.18em K}}
\def\IL{\relax{\rm I\kern-.18em L}}
\def\IM{\relax{\rm
I\kern-.18em M\kern -.18em I}}

\def\IR{{\bf R}}
\def\IZ{{\bf Z}}

\def\CN{{\cal N}}

\def\CR{{\cal R}}


\def\p{\partial}

\def\ib {\bar i}
\def\jb {\bar j}


\def\ib{{\bar i}}
\def\Tr{\rm Tr}

\def\MM{\widehat{\cal M}}

\font\manual=manfnt \def\dbend{\lower3.5pt\hbox{\manual\char127}}

\def\inbar{\,\vrule height1.5ex width.4pt depth0pt}



\def\boxit#1{\vbox{\hrule\hbox{\vrule\kern8pt
\vbox{\hbox{\kern8pt}\hbox{\vbox{#1}}\hbox{\kern8pt}}
\kern8pt\vrule}\hrule}}
\def\mathboxit#1{\vbox{\hrule\hbox{\vrule\kern8pt\vbox{\kern8pt
\hbox{$\displaystyle #1$}\kern8pt}\kern8pt\vrule}\hrule}}


\def\inbar{\,\vrule height1.5ex width.4pt depth0pt}

%

\def\a{\alpha}
\def\b{\beta}
\def\ad{\dot\alpha}

\def\g{\gamma}
\def\e{\epsilon}
\def\ve{\varepsilon}
\def\m{\mu}

\def\l{\lambda}

\def\lb{\bar\l}
\def\ps{\psi}
\def\psb{\bar\psi}
\def\mm{{\bf m}}
\lref\bss{L.~Brink, J.H.~Schwarz, J.~Sherk, Nucl. Phys. {\bf B} 121 (1977) 77}
\lref\witdyn{E. Witten, ``String theory dynamics
in various dimensions,''
hep-th/9503124, Nucl. Phys. {\bf B} 443 (1995) 85-126}
\lref\witbound{E.~Witten, ``Bound States Of Strings And $p$-Branes'',
hep-th/9510135
Nucl. Phys. {\bf B}460 (1996) 335-350}
\lref\vw{C.~Vafa, E.~Witten, ``A strong coupling test of
$S$-duality'', hep-th/9408074;
Nucl. Phys. {\bf B} 431 (1994) 3-77}
\lref\bfss{T.~Banks, W.~Fischler, S.~Shenker, L.~Sussking, Rhys. Rev. 
{\bf D}55 (1997) 5112-5128}
\lref\mns{G.~Moore, N.~Nekrasov, S.~Shatashvili, 
``D-particle bound states and generalized instantons'', 
hep-th/9803265}
\lref\grn{M.~Green, M.~Gutperle, ``D-particle bound states and the
$D$-instanton measure'',
hep-th/9711107}
\lref\asym{M.~Halpern, C.~Schwartz, Int. J.Mod. Phys. {\bf A}13
(1998) 4367-4408} 
\lref\sav{S.~Sethi, M.~Stern, ``$D$-brane bound states redux,''
hep-th/9705046}
\lref\pyi{P.~Yi,
``Witten Index and Threshold Bound States of D-Branes''
hep-th/9704098,  Nucl. Phys. {\bf B} 505 (1997) 307-318}
\lref\higgs{G.~Moore, N.~Nekrasov, S.~Shatashvili, ``Integrating over
Higgs Branches'',
hep-th/9712241}
\lref\kato{S. Hirano, M. Kato, ``Topological Matrix Model',  
hep-th/9708039, Prog.Theor.Phys. 98 (1997) 1371.}
\lref\horostrom{G.T. Horowitz and A. Strominger,
``Black strings and $p$-branes,''
Nucl. Phys. {\bf B360}(1991) 197}
\lref\nicolai{W.~Krauth, H.~Nicolai, M.~Staudacher,
``Monte Carlo Approach to M-theory'', hep-th/9803117}
\lref\manin{Yu.~Manin, ``Generating functions in algebraic
geometry and sums over trees'', alg-geom/9407005}
\lref\estring{J.A.~Minahan, D.~Nemeschansky, C.~Vafa, N. P.~Warner,
``E-Strings and $\CN=4$ Topological Yang-Mills Theories'',
hep-th/9802168}
\lref\townsend{P. Townsend, ``The eleven dimensional supermembrane
revisited,'' hep-th/9501068}
\lref\porrati{M.~Porrati, A.~Rozenberg,
``Bound States at Threshold in Supersymmetric Quantum Mechanics'',
hep-th/9708119}
\lref\ikkt{N. Ishibashi, H. Kawai, Y. Kitazawa, and A. Tsuchiya,
``A large $N$ reduced model as superstring,'' hep-th/9612115;
Nucl. Phys. {\bf B498}(1997)467.}
\lref\sasha{A.~Turbiner, Comm. Math. Phys.  {\bf 118} (1988), 467}
\lref\polch{J.~Polchinski, ``M-Theory and the Light Cone'', hep-th/9903165}
\lref\sussk{L.~Susskind, 
``Holography in the Flat Space Limit'', hep-th/9901079}
\lref\landau{L.~D.~Landau, E.~M.~Lifshitz, 
``Quantum Mechanics'', Pergamon Press, 1977}
\lref\marty{M.~Claudson, M.B.~Halpern, `Supersymmetric ground state wave
functions'', 
Nucl. Phys. {\bf B} 250 (1985)  689}
\lref\polchin{S. Chaudhuri, C. Johnson, and J. Polchinski,
``Notes on D-branes,'' hep-th/9602052; J. Polchinski,
``TASI Lectures on D-branes,'' hep-th/9611050}
\lref\hoppe{G.~Graf, J.~Hoppe, hep-th/9805080}
\lref\hoppei{M.~Bordemann, J.~Hoppe, R.~Suter, hep-th/9909191}
\lref\smilga{V.~Kac, A.~Smilga, hep-th/9908096}
\Title{ \vbox{\baselineskip12pt\hbox{hep-th/9909213}
\hbox{PUPT-1902}
\hbox{ITEP-TH-8/99}
\hbox{NSF-ITP-99-28}}}
{\vbox{
\centerline{On The Size of The Graviton}
}}
\bigskip
\bigskip
\centerline{Nikita A.~Nekrasov}

\bigskip
\centerline{\cyr Institut Teoretiqesko\u\i\quad i E1ksperimentalp1no\u\i\quad Fiziki, 
117259, Moskva, Rossiya}
\centerline{\rm Joseph Henry Laboratory, Princeton University, Princeton NJ 
08544}
\centerline{\rm Institute of Theoretical Physics, 
University of California, Santa Barbara CA 93106 USA}

\medskip
\centerline{\tt  nikita@bohr.harvard.edu}
\bigskip\bigskip\bigskip
We propose an approximate  
wavefunction of the bound state of $N$ $D0$-branes. Its 
spread grows as $N^{1\over 3}$ per particle, i.e. it saturates
the Polchinski's bound.

\vskip 2cm
{\tt Contribution to the Proceedings of the QFTHEP'99} 
\Date{10/99}

\newsec{Introduction}

Definition of M-theory \townsend\witdyn\ 
using its ten dimensional compactification
requires  the knowledge of the behaviour
of the Kaluza-Klein modes of the graviton. These are realized as the bound
states \witbound\ 
of $D0$-branes of Type IIA string theory, originally discovered
as the black holes in the supergravity \horostrom. The aim of this paper
is to get a good grip on the wavefunction $\Psi_{N}$ of the boundstate of
$N$ $D0$-branes, for any $N$. 

Of course, the quantum mechanics of $N$ $D0$-branes contains sectors 
with $N$ free particles, with several clumps of $D0$-branes
bound to each other. In other words, for each partition $\vec n$:
$$
N = 1 \cdot n_1 + 2 \cdot n_2 + \ldots + k \cdot n_k
$$
there is a state whose wavefunction    
looks like:
$$
\Psi_{\vec n} \sim \prod_{l=1}^{k} \Psi_{l} ( x_{l,1} )
\Psi_{l} ( x_{l,2}) \ldots \Psi_{l} (x_{l, n_{l}} ) 
e^{i p_{l, 1} \cdot x_{l,1} + \ldots i p_{l, n_{l}} \cdot x_{l, n_{l}}}
$$

Our approximation contains all such states, which allows
to compute the decay rates and overlaps between multiparticle
states and single-particle states.   

We are not going to 
review the arguments in favor of the existence of the bound state
(see, e.g. \bfss\porrati\grn\sav\pyi\mns\asym).
We proceed with the  variational approach to the 
problem of finding of the ground state. To do this we perturb the
quantum mechanical problem: $ {\hat H}_{0} \to {\hat H}_{\mm}$. The
new problem with the Hamiltonian ${\hat H}_{\mm}$ has only 
four supercharges
instead of the sixteen which were the symmetries of ${\hat H}_{0}$.
But the advantage is the better control on the spectrum
of ${\hat H}_{\mm}$. We
then take the approximate ground state of the perturbed model as the trial 
 wavefunction for the original problem and minimize the
bosonic contribution to the
energy with respect to the parameter $\mm$ of the perturbation. 
Of course the standard approach would be to minimize
the full energy. But
our trial wavefunctions are the ground states of the perturbed
Hamitlonian  so that we need to find another
optimizing criterium.
We suggest to look at the bosonic
potential which is for large $N$ parametrically
equivalent to looking at the spread
$\langle {\Tr} X^2 \rangle$ of the ground state. 
In this way we get 
for the optimal value of 
the parameter $\vert \mm  \vert \sim N^{-{2\over 3}}$.

As a general remark we would like to 
stress that our approach cannot be viewed as a completely satisfactory
one, but nevertheless we consider it useful as it allows
to study the large $N$ case and serves as a nice complement
to the asymptotic approach of \asym\ and subsequent work \hoppe\
(see \hoppei\ for the further apology for our point
of view). Also it seems very simple to generalise our approach
for the other gauge groups, rather then $SU(N)$
thus extending the work \smilga.

As one of the most interesting
applications we suggest to  
compute the overlap between two-particle
wavefunction with particles of mass $N_{1}, N_{2}$ 
and  the single state with $N = N_{1} + N_{2}$. We hope
to return to this problem in the near future.

\noindent 
{\bf Acknowledgements.} I would like to thank J.~Polchinski for discussions.
Research was supported  by Harvard Society of Fellows,
  partly by NSF under the
grants
PHY-94-07194,
  PHY-98-02709, partly by
RFFI under grant 98-01-00327,
partly  by the grant
96-15-96455 for scientific schools. I am grateful to ITP, Santa Barbara
and to ESI, Vienna  
for hospitality.

\newsec{Quantum mechanics}

Consider the quantum mechanics of $D0$-branes in Type $\II$A theory
in the flat Minkovski space. As it is well-known \witbound\ 
in the sector
with $N$ particles they are described by the dimensional reduction of the
$U(N)$ 
$\CN=4$ super-Yang-Mills 
theory down to $0+1$ dimensions (such gauge quantum mechanics were actually introduced a long time ago \bss\marty). 
Upon excluding the center-of-mass motion it becomes $SU(N)/{\IZ}_{N}$ gauge
quantum mechanics.

\noindent
{\bf Hamiltonian and Symmetries.} The  Hamiltonian of the model operates
on the spinor wavefunctions. It is given by:
\eqn\hamil{{1\over 2} {\Tr} P_{M} P_{M} - {1\over 4} {\Tr} [ X^{M}, 
X^{N} ]^2 - {1\over 2} {\Tr} \psi \Gamma_{M} [X^{M}, \psi]}
where all fields are traceless hermitian matrices, the indices $M, N$
run from $1$ to $9$, repeated indices are summed over,
in addition, 
$\psi$ are the Majorana-Weyl fermions of $SO(9)$, 
$\Gamma_{M}$ are the Dirac matrices of $SO(9)$.

The canonical commutation relations are:
\eqn\cancom{[ P_{M, A}, X^{N}_{B} ] = \delta^{N}_{M}\delta_{AB}, 
\quad \{ \psi_{\a, A}, 
\psi_{\b, B} \}  = \delta_{\a\b} \delta_{AB}
}
where $A, B$ are the $SU(N)$ indices.

The supersymmetry of this model is generated by the sixteen 
supercharges:
\eqn\supchrg{Q_{\a} = \Gamma_{\a\b}^{M} {\Tr} P_{M} \psi_{\b} - {i\over 4}
[\Gamma^{M}, \Gamma^{N}]_{\a\b} {\Tr} \psi_{\b} [X^{M}, X^{N}] }

It is sometimes convenient to think of this supersymmetric model as of
the reduction of the four dimensional $\CN=4$ super Yang-Mills theory. 
In this formulation only the $S0(3)_{E} \times SO(6)_{R}$ part of  
the global symmetry group $SO(9)$ (or more presicely the
cover of the that) is manifest. In our approach we break 
this symmetry. Namely we want to work in the $\CN=1$ 
four dimensional terms.  The choice of
$\CN=1$ subalgebra breaks $SO(6)_{R}$ down to $SU(3)_{R}$.
  
In the $\CN=1$ four dimensional terms the field  content of the 
problem is that of a vector multiplet $\left( A_{\m}, \l_{\a} \right)$ 
and a triple of chiral multiplets
$\left( q^{i}, \psi^{i}_{\a} \right)$, 
with the superpotential 
\eqn\supr{W_{0} = {i\over 6} \epsilon_{ijk} {\Tr} q^{i} [ q^{j}, q^{k}]}
The indices $\m$ run from $0$ to $3$, 
$i = 1,2,3$ is the index in the ${\bf 3}$ of $SU(3)_{R}$, $\a = 1,2$
is the index in the $\bf 2$ of the spin cover of $SO(3)_{E}$.  
The classical vacua are the minima of the associated potential
\eqn\pot{V = \sum_{i} {\Tr} F_{i} F_{\ib} +
{\Tr} D^2 + \sum_{m, i} {\Tr} [A_{m}, q^{i}][q^{i, \dagger}, A_{m}]
- \sum_{m < n} {\Tr} [A_{m}, A_{n}]^2,}
$m,n = 1,2,3$, where 
\eqn\fdtrms{\eqalign{F_{i} = {{\p W_{0}}\over{\p q^{i}}},  & \quad F_{\ib} = F_{i}^{\dagger}, \cr
D =
{1\over 2} \quad & \sum_{i} [ q^{i}, q^{i, \dagger}]\cr }}
The vector multiplet fields transform in $({\bf 3}, {\bf 1}) \oplus
({\bf 2}, {\bf 1})$ of $SO(3)_{E} \times SU(3)_{R}$. The chiral multiplets
fall in $({\bf 1}, {\bf 3}) \oplus ({\bf 2}, {\bf 3})$.

\noindent
{\bf Deformation.}
We are going to study the deformed quantum mechanical model, where
the superpotential is replaced by:
\eqn\supri{W_{\mm} = W_{0} + {1\over 2} {\mm} {\Tr} q^{i} q^{i} }
This deformation breaks the $SO(9)$ global symmetry down to 
$SO(3)_{E} \times SO(3)_{R}$, where $SO(3)_{R}$ is imbedded into $SU(3)_{R}$
as a subgroup of unitary transformations preserving the 
`metrics' $\delta_{ij}$. The parameter of the deformation is the complex number
$\mm$.

The deformation \supri\ 
was succesfully used in four dimensions in the analysis
of the topologically twisted $\CN=4$ theory \vw, in the problem of
evaluating the index of the quantum mechanical problem \mns, 
in the arguments in favor of the existence  of the bound state  \porrati\ 
(for prime $N$).

The critical points of the deformed superpotential are the solutions to:
\eqn\vceq{[q^{j}, q^{k} ] = i{\mm} \ve_{ijk} q^{i}}
i.e. ${1\over{\mm}} q^{k}$ must form a representation of the $sl_{2}$ Lie
algebra. All finite-dimensional representations of
$sl_{2}$ are unitary and decompose into sums of the irreducible 
representations:
\eqn\splt{{\bf N} = {\bf 1} \cdot v_1 \oplus {\bf 2} \cdot v_2 \ldots
\oplus {\bf k} \cdot v_{k},}
$v_{l}$ are the multiplicities.
The minima of the potential \pot\ must in addition obey 
\eqn\minpot{[A_{m}, q^{i}  ] = 0, m = 1,2,3, \quad [ A_{m}, A_{n} ] = 0}
which implies that $A_{m}$ must belong to the Lie 
algebra of 
\eqn\unbrg{H = S\left( U(v_1) \times \ldots U(v_{k})\right).}
Although for generic choice of $v_{i}$'s the only massless
modes are those of  gauge fields taking values in $H$, 
for the special choices of $v_{i}$ one gets extra matter.

Let $V$ be the original complex $N$-dimensional
space (the space of Chan-Paton indices), $V_{k} = {\IC}^{v_{k}}$
the multiplicity space, 
${\CR}_{k} \approx {\IC}^{k}$ the standard spin ${{k-1}\over{2}}$
representation  of $sl_{2}$. Then \splt\ can be rewritten as
equality of two reprsentations of $sl_{2}$:
\eqn\splti{V = \bigoplus_{k} V_{k} \otimes {\CR}_{k}}
Due to pseudo-reality of the representations of $SU(2)$ we have
$V \approx V^{\dagger}$, although this isomorphism goes through
the non-trivial transformation
$L_{i} \to - L_{i}^{\dagger}$.
Now all the fields of our quantum mechanical system can be written according
to their $sl_2$ transformation properties.

\item{} The gauge field $A_{m} \in V \otimes V \otimes {\CR}_{1} \otimes {\IR}^3$

\item{} The fermions $\psi^{i}_{\a} \in V \otimes V \otimes {\CR}_{3} \otimes
{\IC}^{2}$

\item{} The fermions $\l_{\a} \in V \otimes V \otimes {\CR}_{1} \otimes {\IC}^{2}$

\item{} The scalars $q^{i} \in V \otimes V \otimes {\CR}_{1}$

\noindent
The importance of this representation is justified by the following
statement:
\lref\lnv{A.~Lawrence, N.~Nekrasov and C.~Vafa,  hep-th/9803015}
\lref\ksorb{S. Kachru and E. Silverstein, hep-th/9802183}
{\it In the quadratic approximation the Hamiltonian of the model is equal to:}
\eqn\quadham{H_{2} = \sum_{\rm fields} {\Phi}^{\dagger} C_{2} {\Phi},
\quad C_{2} = \sum_{i} L_{i} L_{i}}
The tensor product $V \otimes V$ always contains at least one component
of spin zero. This is the trace part of the matrices which 
must be projected out for we deal with $SU(N)$ rather then $U(N)$ fields.
Of course, if more then one $v_{k}$ is different from zero then 
the spin zero components in the product $V \otimes V$ survive. 
In the problem of the computation of the 
index such choices of $\vec v$ contribute zero (see \mns\ for the principal
contribution and \grn\ for the boundary terms). 

\noindent  
It is perhaps worthwhile mentioning 
 here that the set of massless modes coincide
with the field content of the theory obtained by orbifolding of the
$\CN=4$ super-Yang-Mills theory by the $SO(3)$ subgroup of $SU(3)$
in the spirit of \ksorb\lnv\foot{The idea to orbifold by the infinite
subgroups of $SO(6)$ arose in the discussion with C.~Vafa and was
dismissed by both of us as the crazy one. It still looks like
the one but may prove to be useful. }

The next section is devoted to the justification of the claim about the
quadratic Casimir. 
Let $L_{i}$ be the generators of $SU(2)$ in the representation $V$,
\eqn\sutwo{[L_{i}, L_{j} ] = i \ve_{ijk} L_{k}}
 
\noindent{\bf Bosonic potential.}
Write:
\eqn\nota{\eqalign{
q_{i} = {\mm} \left( L_{i} + \xi_{i} \right),  &
\quad q_{i}^{\dagger} = {\bar\mm} \left( L_{i} + \xi_{i}^{\dagger} \right), \cr
\xi_{i} = \b_{i} + i \g_{i},  & \quad A_{m} = \vert {\mm} \vert \a_{m} \cr}}
with $\a_{m}, \b_{i}, \g_{i}$ being Hermitian matrices.
A little computation shows that  
the effective bosonic potential  looks like:
\eqn\effpoti{\eqalign{
{1\over{\vert {\mm} \vert^4}} V_{\xi} = {1\over 2} \sum_{m}
{\Tr} \a_{m} C \a_{m}  +  & {1\over 2} \sum_{j} 
{\Tr} \left( \b_{j} ( C + 2)  \b_{j} +   \g_{j} (  C +  2) \g_{j} \right)  \cr
- \sum_{i,j,k} 
{\Tr} \ve_{ijk} L_{k} \left( [ \b_j, \b_i ] + [\g_j, \g_i] \right) + & 
 {\Tr} {\tilde D}^2 \cr
\tilde D \quad = & \sum_{j} [ L_j, \b_j] \cr
C K = \quad & \sum_{j} [L_{j}, [L_{j}, K]]\cr }}

\noindent{\bf Bosonic eigenmodes.}  
The bosonic potential \effpoti\ gives masses to nine bosonic
traceless hermitian matrices, provided that the gauge fixing is
done appropriately. The point is that  the critical points
of the superpotential $W_{\mm}$ form a continious family,
due to the invariance of  \vceq\ under the complexified
gauge transformations from ${\rm SL}_{N}({\IC})$. 
The non-compact part is broken by the presence of the 
square of the moment map term ${\Tr} \sum_{i}\left( [ q_{i} ,
q_{i}^{\dagger}]\right)^2$ but the compact part is not.
The choice of the representatives $L_i$ breaks the 
$SU(N)$ invariance but by the well-known Goldstone effect
the gauge group reveals itself in the presence of the
massless modes, of the form:
\eqn\msls{\b_j = i [ L_j, \chi ], \quad \chi^{\dagger} = \chi }
These massless modes are eventually killed by the Gauss law, so we might 
as well work in the gauge:
\eqn\gge{{\tilde D} = 0}
To see whether \gge\ is a good gauge choice let us make an infinitesimal
gauge transformation  {\it \`a la} \msls. We see that
$\tilde D$ changes by $i C \chi$. If the traceless part of
$V \otimes V$ does not contain spin zero pieces then $C$ is invertible and
\gge\ is a good gauge choice. This is equivalent to 
the condition $v_{k} = \delta_{k, N}$, which is the case of our
utmost interest.

\noindent{\bf Fermionic potential.}
The fermionic part of the Hamiltonian can be written in the
following way (the indices $\a,\b$ are raised and lowered
with the help of $\e_{\a\b} = - \e_{\b\a}$ symbol, the indices
$i,j,k$ are raised with the help of $\delta_{ij}$,
we also have a $\delta_{i\ib}$ pairing):
\eqn\fer{\eqalign{& {1\over{2\sqrt{2}}} {\Tr} \left( 
i \ve_{ijk}   q^{i} [ \ps^{j}_{\a}, \ps^{k, \a} ] 
+ \mm \ps_{i, \a} \ps^{i, \a}  
+ q^{\ib} [\ps^{i}_{\a}, \l^{\a}]  \right) - \cr
& - {1\over{2\sqrt{2}}} {\Tr} \left( 
i \ve_{\ib\jb\bar k}   q^{\ib} [ \psb^{\jb}_{\ad}, \ps^{\bar k, \ad} ] 
+ {\bar\mm} \psb_{\ib, \ad} \psb^{\ib, \ad}  
+ q^{i} [\psb^{\ib}_{\ad}, \lb^{\ad}]  \right)  \cr}
}

In the quadratic approximation it reduces to:
\eqn\qfer{{{\mm}\over{2\sqrt{2}}} {\Tr} \left( i
L_{i} [\ps^{j}_{\a}, \ps^{k,\a}] \ve_{ijk}  + \ps^{i}_{\a} \ps^{i,\a} -
i L_{i} [\psb_{\ib,\ad}, \lb^{\ad} ] \right) + {\rm c.c.}}

\noindent
{\bf Completely Higgsed phase.}
The case of our immediate interest is $v_k = \delta_{k, N}$, i.e. 
when the representation is irreducible. In this case the solution to
\minpot\ is $A_{m} = 0$ i.e. all fluctuations of  $A_{m}$'s are massive.
In addition, since the irreps of $SU(2)$ have no moduli the fluctuations
of $q_{i}$ which are orthogonal to the gauge orbit
 are massive as well. Let $L_i$ denote the standard
generators of $SU(2)$ in the $N$-dimensional representaion:
\eqn\repr{\eqalign{L_{3} = & {\rm diag} \left( -j, \ldots, j \right) \cr 
L_{\pm} = &   L_{1} \pm i L_{2}  \cr
\left[ L_{\pm} \right]_{m\pm 1, m} = & \sqrt{(j+1 \pm m) (j \mp m)} \cr }}
Then at the minimum of the potential 
$q_{i}  = {\mm} L_{i}, q_{\ib}^{\dagger} = {\bar\mm} L_{i}$, $A_{m} = 0$. 

\newsec{The size of the bound state}

The zero energy eigen-function of the deformed Hamiltonian  
can be approximated by the Gaussian wavefunction, which is
annihilated by the quadratic Hamiltonian \quadham\ and is localised
near the minimum of the potential given by the formulae \repr.
Let us estimate the bosonic spread of this wavefunction.

We want to evaluate
\eqn\vev{\Delta = \left\langle \sum_{M} {\Tr} X^{M} X^{M} \right\rangle =}
$$
\sum_{m} \langle {\Tr}A_{m}^2 \rangle + \sum_{i} \langle 
{\Tr} q_{i}q_{\ib}^{\dagger} \rangle =
\vert \mm \vert^2 \sum_i {\Tr} L_i^2 + 
{3\over{2\vert \mm \vert}}  
{\Tr}_{V \otimes V - {\CR}_{0}}
{1\over{C^{1\over 2}}}  + 
{1\over{\vert \mm \vert}}  
{\Tr}_{V \otimes V \otimes {\CR}_1 - {\CR}_{0}}
{1\over{C^{1\over 2}}} =
$$
$$
={3\over 4}\vert \mm \vert^2    N (N^2 -1)+
{9 \over {2\vert \mm \vert}} \sum_{l=1}^{N-2} {{2l+1}\over{\sqrt{l(l+1)}}}
+ 4{{2N-1}\over{\sqrt{N(N-1)}}} + 2{{2N+1}\over{\sqrt{N(N+1)}}}   
$$
For $N$ large $\Delta$ goes like:
\eqn\lrgn{{\Delta} \sim {3\over 4} \vert \mm \vert^2 N^3 +
9 {N\over{\vert \mm \vert}}}
Let us now adjust  $\mm$ so as to minimize $\Delta$.
We expect to have small corrections both to the 
shape of the wavefunction and its spread for this value of the
parameter $\mm$ coming from the non-linear terms in the 
original Hamiltonian as well as from the added terms.
We get:
$$
{\mm} \sim {6^{1\over 3}   \over N^{2\over 3}},
\qquad
{\Delta} \sim 7.43 N^{5\over 3},
$$
in agreement with the estimates \polch\sussk.

\appendix{A}{\bf  Diagonalization of the effective mass matrix.}

We want to diagonalize the effective mass operator:
\eqn\effm{{\MM} : \phi \mapsto \sum_{i} [L_{i}, [L_{i}, \phi]] }
In the basis where the generators of $sl_{2}$ are represented as
\repr\ the eigen-value problem for the operator \effm\ 
${\MM} \phi = E \phi$ is written explicitly as:
\eqn\eigen{\eqalign{ \sqrt{(j+1 + m^{\prime}) ( j - m^{\prime} ) ( j + 1 + m )(j-m)} \phi_{m+1, 
m^{\prime} +1} & 
+ 2 m m^{\prime} \phi_{m m^{\prime}} +  \cr
+ \sqrt{( j + m^{\prime} )(j+m)( j+1-m)(j+1 - m^{\prime})}
 \phi_{m-1, m^{\prime}-1} = 
&  \left( 2  j(j+1) - E \right)
\phi_{m m^{\prime}} \cr}}

\noindent
{\bf Reduction to the representation theory problem.}
Let $V$ denote the $N$-dimensional representation of $su(2)$, so 
$\phi \in {\rm Hom} (V, V)$. 
In solving the eigen-value problem \eigen\ we are allowed to perform
the transformations $\phi \to g \phi h^{-1}$
where $g, h$ are the group elements of  $SU(2)$ acting in 
$V$. By applying such transformations we can map the operator
$\MM$ to the operator $\bf M$ acting in the space $V \otimes V$:
\eqn\opr{{\bf M} = \sum_{i=1}^{3} \left(
L_{i}^{2} \otimes 1 +  1 \otimes L_{i}^2 + 2 L_{i} \otimes L_{i} \right)}
Let ${\hat L}_{i} = L_{i} \otimes 1 + 1 \otimes L_{i}$ be the
generator $L_{i}$ acting in $V \otimes V$. Then 
\eqn\oppr{{\bf M} =  \sum_{i} {\hat L}_{i}^2,}
i.e. it is simply the quadratic Casimir acting in the tensor product. 
Clearly, $\bf M$ is diagonalized by decomposing $V\otimes V$ into irreducibles:
\eqn\dcmp{V \otimes V = \bigoplus_{l=0}^{N-1} {\CR}_{l}, \quad {\bf M} =
\oplus_{l=0}^{N-1}  l (l+1)  {\bf 1}_{2l+1} } 
where ${\CR}_{l}$ is the spin $l$ irrep of $SU(2)$.

The zero mode corresponding to $l=0$ is the center-of-mass
degree of freedom, $\phi_0 \sim {\bf 1}_{N}$ and is projected out
by the condition that ${\Tr} \phi = 0$. 

\noindent
{\bf Polynomial representation I.}
Let us form the  generating function:
\eqn\gnfn{\Phi (x, y) = \sum_{m, m^{\prime}}
{{\phi_{m m^{\prime} } x^{j+m} y^{j+m^{\prime}}}\over{
\sqrt{(j+m)! (j - m)! ( j - m^{\prime})! (j + m^{\prime})!}}}}
Then the operator $\MM$ is represented by the first order 
bi-differential operator:
\eqn\diffrp{{\MM} = 2j(2j + 1 )  - ( xy + 1)^2 \p_{x} \p_{y} + 
2j ( 1+ xy) \left( x\p_{x} + y\p_{y} - 2j \right)}
It is natural to pass to the coordinates $y, \rho = xy$. Then the
solution to the eigen-value problem can be found in the separated form:
\eqn\asol{\Phi (x, y) = y^{a} \psi_{a} ({\rho})}
where for positive $a$ the function $\psi_{a}({\rho})$ is
a polynomial of degree $2 j  - a$, while for negative
$a = - b < 0$ the function $\psi_{a} $ has to be of the form
$\psi_{a} ( \rho ) = {\rho}^{b} {\tilde\psi}_{b} ({\rho})$ where
${\tilde\psi}_{b}({\rho})$ is a polynomial of degree $2j - b$.
The function $\psi_{a}$ is an eigen-function of the second-order
differential operator:
\eqn\sepa{2j - \left( \rho + 1 \right)^2 \left( 
\rho \p_{\rho}^2 + \left( a + 1 \right) \p_{\rho} \right) +
2j \left( \rho + 1 \right) \left( 
2\rho \p_{\rho} + a \right) - 4 j^2 \rho
}    
Let $t = 1 + \rho$. 
For $a \geq 0$ 
introduce generators of $sl_2$ (following the general method of
\sasha):
\eqn\sltwo{\eqalign{\ell_{+} \quad = & \quad - t^2 \p_{t} + ( 2j - a) t\cr
\ell_{0} \quad = & \quad t \p_{t} - j + {a\over 2} \cr
\ell_{-} \quad = & \quad \p_{t} \cr}} 
Then the operator \sepa\ can be written as:
\eqn\sepai{ \left( \ell_{+} + \ell_{0} - j - {{a}\over{2}} - 1 \right)
\left( \ell_{0} - j - {{a}\over{2}} \right)}
which is an upper-triangular matrix in the basis 
$1, t, \ldots, t^{2j-a}$ of the monomials, which form an invariant subspace for
the operator \sepa\ in the space of all polynomials (such operators are
called quasi-exactly-solvable \sasha) . Its diagonal entries
are the eigen-values, hence the spectrum of $\MM$ is given by:
\eqn\spectrum{E_{n,a}  = (N - n) ( N - n - 1),
\quad n = 0, \ldots, N - 1 - a, \quad a = 0, \ldots, N-1 }
For $a < 0$ the spectrum is identical. 
  
\noindent
{\bf Polynomial representation $\II$.} Let us form another
 generating function:
\eqn\gnfnii{{\tilde\Phi} (x, y) = \sum_{m, m^{\prime}}
{{\phi_{m m^{\prime} } (-x)^{j+m} y^{j- m^{\prime}}}\over{
\sqrt{(j+m)! (j - m)! ( j - m^{\prime})! (j + m^{\prime})!}}}}
Then the operator $\MM$ is represented by the first order 
bi-differential operator:
\eqn\diffrp{{\MM} = 2j(2j + 1 )  - ( x - y)^2 \p_{x} \p_{y} -
2j ( x- y) \left( \p_{x} - \p_{y}  \right)}
It is natural to pass to the coordinates $t = {{x-y}\over{x+y}}, 
\rho = x + y$. Then the
solution to the eigen-value problem can be found in the separated form:
\eqn\asol{{\tilde\Phi} (x, y) = \rho^{a} \psi_{a} (t), \quad, 0 \leq a \leq 
4j }
where the function $\psi_{a}(t)$ is
a polynomial of degree $a$. It is also 
 an eigen-function of the second-order
differential operator:
\eqn\sepaii{- {\ell}_{+}^2 + \left( {\ell}_{0} + {a\over 2} - 2j \right)
\left( {\ell}_{0} + {a\over 2} - 2j - 1 \right)}
where we have
introduced generators of $sl_2$:
\eqn\sltwo{\eqalign{\ell_{+} \quad = & \quad - t^2 \p_{t} + a t\cr
\ell_{0} \quad = & \quad t \p_{t} -  {a\over 2} \cr
\ell_{-} \quad = & \quad \p_{t} \cr}}
which make \sepaii\  an upper-triangular matrix in the basis 
$1, t, \ldots, t^{a}$ of the monomials. Its diagonal entries
are the eigen-values, hence the spectrum of $\MM$ is given by:
\eqn\spectrum{E_{n,a}  = (N - n) ( N - n - 1),
\quad n = 0, \ldots, a, \quad a = 0, \ldots, 2N-2 }
For $a < 0$ the spectrum is identical. 

\noindent
{\bf Clebsch-Gordan coefficients.} For reader's convenience we list here
the relevant $3j$-symbols, quoting them from \landau:
\eqn\cg{\eqalign{\pmatrix{ j & j & l \cr m & - n & n - m \cr} = & \left( 
{{
(j + m)! (j - m)! (j + n )! (j - n)! (l + m -n )! (l+n-m)!}\over{
(2j - l)! (l!)^2 (2j + l + 1)!}} \right)^{1\over 2} \times \cr
& \sum_{k=0}^{2j - l} (-1)^{k + m + n}
\pmatrix{ l \cr j - m - k\cr} \pmatrix{ l \cr
j - n- k\cr} \pmatrix{ 2j - l \cr k \cr}\cr} }

\listrefs
\bye